\documentclass[10pt,twocolumn,twoside]{IEEEtran}
\usepackage[utf8]{inputenc}         
\usepackage[english]{babel}
\usepackage{multirow}

\usepackage{multicol}

\usepackage{soul}

\usepackage[normalem]{ulem}

\usepackage{xcolor,colortbl}

\ifCLASSINFOpdf
\usepackage[pdftex]{graphicx}
\usepackage{epstopdf}
\graphicspath{{./image/}}
\else
\fi
\usepackage[cmex10]{amsmath}
\usepackage{amssymb}
\usepackage{array}
\usepackage{rotating}

\DeclareMathOperator*{\argmax}{argmax}

\usepackage{makecell}
\usepackage{subcaption}
\usepackage{textcomp}

\definecolor{LightBlue}{rgb}{0.6172,0.7891,0.8789}

\usepackage{lipsum}

\usepackage{mdframed}

\usepackage{url}

\usepackage{cite}
\usepackage{capt-of}

\usepackage{tabularx}
\usepackage{bm}
\usepackage{calc}

\usepackage{xcolor,colortbl}
\usepackage{tabu}
\usepackage{mdframed}

\usepackage{pifont}

\usepackage{balance}

\begin{document}
	%
	%
	%

	\title{Pediatric Automatic Sleep Staging: A comparative study of state-of-the-art deep learning methods}
	\author{{Huy~Phan, Alfred Mertins,
			and~Mathias~Baumert
			\thanks{H. Phan is with the School of Electronic Engineering and Computer Science, Queen Mary University of London, UK and the Alan Turing Institute, UK. A. Mertins is with the Institute for Signal Processing, University of L\"ubeck and the German Research Center for Artificial Intelligence (DFKI), Germany. M. Baumert is with the School of Electronic and Electrical Engineering, University of Adelaide, Australia.}
			\thanks{H. Phan is supported by a Turing Fellowship under the EPSRC grant EP/N510129/1.}
			\thanks{$^*$Corresponding authors: Huy Phan ({\tt\footnotesize h.phan@qmul.ac.uk}), Mathias Baumert ({\tt\footnotesize mathias.baumert@adelaide.edu.au})}
	}}
	
	%
	%
	
	\markboth{THIS ARTICLE HAS BEEN PUBLISHED IN IEEE TRANSACTIONS ON BIOMEDICAL ENGINEERING}%
	{THIS ARTICLE HAS BEEN PUBLISHED IN IEEE TRANSACTIONS ON BIOMEDICAL ENGINEERING}
	%

	\IEEEpubid{DOI: 10.1109/TBME.2022.3174680 \hfill}


	\maketitle
	
	\begin{abstract}
		\emph{Background:} Despite the tremendous progress recently made towards automatic sleep staging in adults, it is currently unknown if the most advanced algorithms generalize to the pediatric population, which displays distinctive characteristics in overnight polysomnography (PSG). 
		\emph{Methods:}	To answer the question, in this work, we conduct a large-scale comparative study on the state-of-the-art deep learning methods for pediatric automatic sleep staging. Six different deep neural networks with diverging features are adopted to evaluate a sample of more than 1,200 children across a wide spectrum of obstructive sleep apnea (OSA) severity. 
		\emph{Results:} Our experimental results show that the individual performance of automated pediatric sleep stagers when evaluated on new subjects is equivalent to the expert-level one reported on adults. Combining the six stagers into ensemble models further boosts the staging accuracy, reaching an overall accuracy of 88.8\%, a Cohen's kappa of 0.852, and a macro F1-score of 85.8\%. At the same time, the ensemble models lead to reduced predictive uncertainty. The results also show that the studied algorithms and their ensembles are robust to concept drift when the training and test data were recorded seven months apart and after clinical intervention.
		\emph{Conclusion:} However, we show that the improvements in the staging performance are not necessarily clinically significant although the ensemble models lead to more favorable clinical measures than the six standalone models. 
		\emph{Significance:} Detailed analyses further demonstrate ``almost perfect'' agreement between the automatic stagers to one another and their similar patterns on the staging errors, suggesting little room for improvement.
	\end{abstract}
	

	%
	\IEEEpeerreviewmaketitle

	\section{Introduction}
	\label{sec:introduction}
	
	Assigning a sleep stage to each 30-second epoch of a full overnight polysomnogram (PSG) is a critical to assess the macro-structure of sleep, i.e., to observe sleep cycles, quantify the time spent in each sleep stage, and determine rapid eye movement (REM) onset latency and wake after sleep onset (WASO). Sleep stages and cycles serve as an important proxy for neuro-physiological processes that orchestrate sleep and provide diagnostic markers for sleep disorders \cite{AASM2014}. For instance, differentiation of sleep stages and sleep stage transitions  are used to quantify sleep continuity in patients with obstructive sleep apnea (OSA) syndrome \cite{Norman2006}, a common sleep disorder in both adults \cite{Senaratna2017} and children \cite{Redline1999}.  Patterns of sleep-stage transitions \cite{Christensen2015} and REM sleep onset \cite{Littner2005} are important indicators for narcolepsy, a rare central hypersomnia. Traditionally, sleep staging has been carried out manually by sleep technicians following the American Academy of Sleep Medicine (AASM) guidelines \cite{Iber2007}. Since manual scoring is time-consuming, laborious, and requires expert knowledge, significant efforts went into teaching a machine to perform sleep staging to reduce costs and make PSG diagnostics more widely available.
	
	The advance of deep learning \cite{LeCun2015} coupled with the establishment of large-scale public sleep databases \cite{Zhang2018, Quan1997} has accelerated automatic sleep staging research. Now, automatic sleep staging systems \cite{Phan2019a,Olesen2021,Phan2021c,Eldele2021,Guillot2021,Phan2021b,Biswal2018a} relying on the new sequence-to-sequence paradigm \cite{Phan2019a} have surpassed the agreement level of experts' scoring \cite{Mikkelsen2021}, reaching an accuracy acceptable for clinical applications. Importantly, this class of sleep staging algorithms has been validated almost exclusively on adult PSGs. How they perform on pediatric PSGs, in particular on clinical populations with OSA, remains uncharted. Evaluating these algorithms on the pediatric PSGs is crucial given their considerable discrepancies to the adult ones. For example, EEG of children shows higher amplitude and a slower dominant posterior rhythm than the alpha rhythm seen in adults \cite{Grigg-Damberger2007}. Furthermore, the sleep architecture in children exhibits significant differences among age subgroups. For example, the amount of REM and slow wave sleep (SWS) changes dramatically during infancy, childhood and adolescence. Also, children tend to move more during sleep than adults, affecting the quality of the recorded data adversely, possibly imposing greater challenges on automated systems. Thus, it is imperative to benchmark established algorithms on the pediatric PSG.
	
	\IEEEpubidadjcol
	In this work, we aim to determine if the expert-level performance of the state-of-the-art sleep scoring algorithms is generalizable to the pediatric population. To this end, we conduct a comparative study to benchmark six different algorithms on a large clinical-validated pediatric cohort of $>\!1,200$ children with $>\!1,600$ PSG recordings in total. All children underwent PSG screening for OSA, displaying a wide range of OSA severity, including negative, mild, moderate, and severe cases. The algorithms adopted in this study adhere to the sequence-to-sequence framework \cite{Phan2019a} and manifest diverging characteristics on their input types and network architectures. They include XSleepNet1 \cite{Phan2021c}, XSleepNet2 \cite{Phan2021c}, SeqSleepNet \cite{Phan2019a}, DeepSleepNet \cite{Supratak2017}, FCNN-RNN \cite{Phan2021c}, and SleepTransformer \cite{Phan2021b}, which were recently reported to achieve state-of-the-art performance on a variety of public adult PSG databases, such as Sleep-EDF Expanded \cite{Kemp2000, Goldberger2000}, Montreal Archive of Sleep Studies (MASS) \cite{Oreilly2014}, Physio2018 \cite{Ghassemi2018}, and Sleep Heart Health Study (SHHS) \cite{Zhang2018, Quan1997}. Furthermore, we propose and evaluate two ensemble models that combine the six adopted sleep stagers via a probability averaging strategy and a convolutional neural network (CNN). Our main contributions are as follows:
	\begin{itemize}
		\item We establish a comprehensive benchmark on a rich set of state-of-the-art deep learning methods for automatic sleep staging on a large-scale pediatric population with a wide range of OSA severity.
		\item We empirically demonstrate that the studied algorithms (i) reach an expert-level accuracy on pediatric sleep staging, similar to that reported for adult PSGs; (ii) are robust to concept drift \cite{Tsymbal2004}; (iii)  agree to one another ``almost perfectly''; and (iv) share similar patterns on their classification errors. 
		\item We show that the proposed ensemble models lead to improved sleep staging performance and reduced predictive uncertainty. 	
		\item We further show that the improvements on sleep staging performance are not necessarily clinically significant.
	\end{itemize}
	
	Few prior works using deep neural networks on pediatric automatic sleep staging such as \cite{Huang2020,Jeon2019} have been reported, however, to the best of our knowledge, this is the first work to comprehensively benchmark state-of-the-art automatic sleep staging algorithms in children using a large-scale clinically relevant dataset.

	\section{Childhood Adenotonsillectomy Trial (CHAT) database}
	\label{sec:database}
	
	We use overnight PSG from the Childhood Adenotonsillectomy Trial (CHAT) \cite{Marcus2013, Redline2011, Zhang2018}, a multi-center, singleblinded, randomized, controlled trial designed to analyze the efficacy of early adenotonsillectomy (eAT) on children. The trial aimed to test whether children randomized to eAT demonstrate greater improvement in cognitive, behavioral, quality-of-life, and sleep measures than children who were randomly assigned to watchful waiting with supportive care (WWSC) \cite{Marcus2013, Liu2017,Liu2018a}. To that end, 464 children, 5 to 9 years of age, with the OSA syndrome were randomly assigned to eAT or the strategy of watchful waiting. Polysomnographic, cognitive, behavioral, and health outcomes were assessed at baseline and at 7 months. Physiological measures of sleep were assessed at baseline and at 7 months with standardized full PSG with central scoring at the Brigham and Women’s Sleep Reading Center. The children were recruited as part of the clinical trial ``Childhood Adenotonsillectomy Study for Children With OSAS (CHAT)'', ClinicalTrials.gov number, NCT00560859.
	
	In total, 1,447 children underwent screening PSGs conducted at different hospitals with various equipment. Children meeting the CHAT inclusion criteria and participating in the trial also had a follow-up PSG.  In our study, only recordings with at least 5 hours of good data (after excluding ``UNKNOWN'' and not zero/near-zero epochs) were used. We formed the following 3 subsets from the database.
	\begin{itemize}
		\item Baseline: 464 children who were randomized to eAT and WWSC. After excluding withdrawn children \cite{Redline2011} and recordings with less than 5 hours of data, 440 recordings were retained.
		\item Follow-up: the same children as in the Baseline subset about 7 months after the intervention (i.e., either eAT and WWSC). 393 recordings were retained after excluding those with less than 5 hours of data.
		\item Non-randomized: 779 children who were screened but were not included in the trial due to, e.g., negative or severe OSA diagnosis. These children are completely different from those in the Baseline and Follow-up subsets. 776 recordings were retained after excluding those with less than 5 hours of data.
	\end{itemize}
	A summary of the subsets is given in Table \ref{tab:databases}. We adopted C4-A1 EEG and ROC-LOC EOG to study single-channel EEG and dual-channel EEG$\cdot$EOG automatic sleep staging. The data, originally recorded at different sampling rates were downsampled to 100 Hz. Segments with zero/near-zero in the recordings due to poor electrode contact were discarded. To deal with different measurement units owing to different equipment, each signal was normalized to the range [-1, 1] by dividing its maximum magnitude. Prior to normalization, values outside 6 standard deviations were clipped. Band-pass filtering with a low cutoff frequency of 0.3 Hz and a high cutoff frequency of 40 Hz was carried out. Finally, the per-recording signal was normalized to zero mean and unit standard deviation. 
	
	\setlength\tabcolsep{1.65pt}
	\begin{table}[!t]
		\caption{Summary of the data subsets formed from the studied database. AHI - apnea-hypopnea index.}
		\vspace{-0.2cm}
		\scriptsize
		\begin{center}
			\begin{tabular}{|>{\centering\arraybackslash}m{0.1in}|>{\arraybackslash}m{0.95in}|>{\centering\arraybackslash}m{0.675in}|>{\centering\arraybackslash}m{0.675in}|>{\centering\arraybackslash}m{0.675in}|>{\centering\arraybackslash}m{0in} @{}m{0pt}@{}}
				\cline{1-5}
				\multicolumn{2}{|c|}{Subset}& Baseline & Follow-up & Non-randomized & \parbox{0pt}{\rule{0pt}{1ex+\baselineskip}} \\ [0.5ex]
				\cline{1-5}
				\multicolumn{2}{|c|}{\#recordings} & 440 & 393 & 776 & \parbox{0pt}{\rule{0pt}{0ex+\baselineskip}} \\ [0.5ex]
				\cline{1-5}
				\multirow{5}{*}{\rotatebox[origin=c]{90}{\#Sleep epochs~~}} & Wake & \makecell{99,764 (19.9\%)} & \makecell{78,528 (17.8\%)}  & \makecell{166,833 (19.3\%)} & \parbox{0pt}{\rule{0pt}{0ex+\baselineskip}} \\ [0.5ex]
				& N1 & \makecell{34,057 (6.8\%)} & \makecell{27,056 (6.2\%)}  & \makecell{55,963 (6.5\%)}  & \parbox{0pt}{\rule{0pt}{0ex+\baselineskip}} \\ [0.5ex]
				& N2 & \makecell{165,531 (33.0\%)}  & \makecell{157,830 (35.9\%)}   & \makecell{294,995 (34.1\%)}  & \parbox{0pt}{\rule{0pt}{0ex+\baselineskip}} \\ [0.5ex]
				& N3 & \makecell{127,643 (25.4\%)}  &  110,554 (25.1\%) & \makecell{218,960 (25.3\%)} & \parbox{0pt}{\rule{0pt}{0ex+\baselineskip}} \\ [0.5ex]
				& REM & \makecell{75,225 (15.0\%)}  & \makecell{66,291 (15.1\%)}  & \makecell{128,216 (14.8\%)} & \parbox{0pt}{\rule{0pt}{0ex+\baselineskip}} \\ [0.5ex]
				\cline{1-5}
				\multicolumn{2}{|c|}{Age} & 6.6$\pm$1.4  & 6.6$\pm$1.4 & 7.1$\pm$1.4 & \parbox{0pt}{\rule{0pt}{0ex+\baselineskip}} \\ [0.5ex]
				\cline{1-5}
				\multirow{5}{*}{\rotatebox[origin=c]{90}{OSA severity~~}} & None (0$\le$AHI$<$1) & 57 & 52 & 464 & \parbox{0pt}{\rule{0pt}{0ex+\baselineskip}} \\ [0.5ex]
				& Mild (1$\le$AHI$<$5) & 223 & 190 &  214 & \parbox{0pt}{\rule{0pt}{0ex+\baselineskip}} \\ [0.5ex]
				& Moderate (5$\le$AHI$<$10)& 106 & 94 & 35 & \parbox{0pt}{\rule{0pt}{0ex+\baselineskip}} \\ [0.5ex]
				& Severe (AHI$\ge$10)& 67 & 57 & 66 & \parbox{0pt}{\rule{0pt}{0ex+\baselineskip}} \\ [0.5ex]
				& Unknown & 11 & 0 & 0 & \parbox{0pt}{\rule{0pt}{0ex+\baselineskip}} \\ [0.5ex]
				\cline{1-5}
				\multirow{4}{*}{\rotatebox[origin=c]{90}{\makecell{Sleep stats~~~}}} & Total sleep time (min)& 456.2 $\pm$ 52.2& 459.0$\pm$ 54.7& 449.3 $\pm$ 56.7  & \parbox{0pt}{\rule{0pt}{0ex+\baselineskip}} \\ [0.5ex]
				& WASO (min) & 46.1$\pm$39.3 & 36.1$\pm$ 31.0 & 42.4 $\pm$ 37.9 & \parbox{0pt}{\rule{0pt}{0ex+\baselineskip}} \\ [0.5ex]
				& REM latency (min) & 226.7$\pm$70.0 & 218.2$\pm$65.7 & 221.3 $\pm$ 66.0 & \parbox{0pt}{\rule{0pt}{0ex+\baselineskip}} \\ [0.5ex]
				& Sleep efficiency (\%) & 81.0$\pm$8.6 & 83.0 $\pm$ 8.2 & 81.8 $\pm$ 9.2 & \parbox{0pt}{\rule{0pt}{-1ex+\baselineskip}} \\ [0.5ex]
				\cline{1-5}
			\end{tabular}
		\end{center}
		\label{tab:databases}
		\vspace{-0.3cm}
	\end{table}
	
	\section{Deep Learning Methods}
	\label{sec:methods}
	
	We adopted a cohort of six different deep neural networks in this study. The networks were chosen taking into account discrepancies in their inputs and network architectures. At the high level, these networks can be fitted neatly into a common framework, namely end-to-end sequence-to-sequence sleep staging framework \cite{Phan2019a,Phan2021a}, which has been the driving force behind expert-level performance in automatic sleep staging reported recently. 
	
\begin{figure} [!t]
		\centering
		\includegraphics[width=0.9\linewidth]{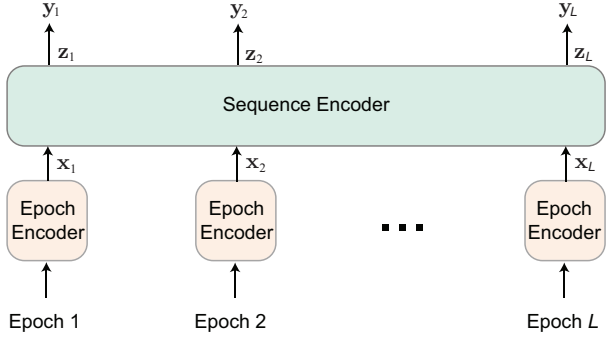}
		\caption{Illustration of the sequence-to-sequence sleep staging framework.}
		\label{fig:seq2seq_framework}
		\vspace{-0.2cm}
	\end{figure}

	Formally, let us denote an input sequence of $L$ epochs as $(\mathbf{S}_1, \ldots, \mathbf{S}_L)$ where $\mathbf{S}_\ell$ is the $\ell$-th epoch, $1 \le \ell \le L$. In general, the epochs can be in any form, such as raw signals or time-frequency images and single- or multi-channel. A network adhering to the framework typically consists of two main components: the epoch encoder $\mathcal{F}_{E}$ and the sequence encoder $\mathcal{F}_{S}$ as illustrated in Figure \ref{fig:seq2seq_framework}. The epoch encoder acts as an epoch-wise feature extractor which transforms an epoch $\mathbf{S}$ in the input sequence into a feature vector $\mathbf{x}$ for representation:
	\begin{align}
		\mathcal{F}_{E}: \mathbf{S} \mapsto \mathbf{x}.
	\end{align}
	As a result, the input sequence is transformed into a sequence of feature vectors $(\mathbf{x}_1, \ldots, \mathbf{x}_L)$. Of note, $\mathcal{F}_{S}$ can be a hard-coded hand-crafted feature extractor, however, in deep learning context, it is oftentimes a neural network (e.g., a convolutional neural network (CNN) or a recurrent neural network (RNN)) that learns the feature presentation $\mathbf{x}$ automatically from low-level inputs. In turn, at the sequence level, the sequence encoder transforms the sequence $(\mathbf{x}_1, \ldots, \mathbf{x}_L)$ into another sequence $(\mathbf{z}_1, \ldots, \mathbf{z}_L)$. Formally, 
	\begin{align}
		\mathcal{F}_{S}: (\mathbf{x}_1, \ldots, \mathbf{x}_L) \mapsto (\mathbf{z}_1, \ldots, \mathbf{z}_L).
	\end{align}
	In intuition, $\mathbf{z}_\ell$ is a richer representation for the $\ell$-th epoch than $\mathbf{x}_\ell$ as it not only encompasses information of the epoch but also interaction with other epochs in the sequence. More specifically, $\mathbf{z}_\ell$ is derived from $\mathbf{x}_\ell$, taking into account the left context $(\mathbf{x}_1, \ldots, \mathbf{x}_{\ell-1})$ and the right context $(\mathbf{x}_{\ell+1}, \ldots, \mathbf{x}_L)$. Eventually, the vectors $\mathbf{z}_1, \ldots, \mathbf{z}_L$ are used for classification purpose to obtain the sequence of predicted sleep stages, one for each epoch in the input sequence.
	
	Below, we concretely describe the adopted networks with respect to their mother framework. A snapshot of the networks is also given in Table \ref{tab:networks}. 
	
	{\bf SeqSleepNet \cite{Phan2019a}}: The network receives a time-frequency representation (i.e., logarithmic magnitude spectrogram) as input. 
	In case of multiple channels, the spectrograms are stacked to form a multi-channel input. On the one hand, the epoch encoder is realized by the coupling of learnable filterbank layers \cite{phan2018c} (one for each input channel), a bidirectional RNN layer, and a gated attention layer \cite{Bahdanau2015,Luong2015b}. On the other hand, the sequence encoder is realized by another bidirectional RNN layer. Both the epoch encoder and sequence encoder have their bidirectional RNN implemented using Long Short-Term Memory (LSTM) cell \cite{Hochreiter1997} coupled with recurrent batch normalization \cite{Cooijmans2016}. 
	
	{\bf SleepTransformer \cite{Phan2021b}}: Similar to SeqSleepNet, SleepTransformer ingests time-frequency input. The network makes use of Transformer \cite{Vaswani2017} as the backbone for both the epoch encoder and sequence encoder, making it distinct from other networks used in this study which are based on either RNN or CNN or both. Leveraging the attention matrices of the Transformers, the network is the first of its kind offering the explainability at both the epoch and sequence level which closely resembles the manual scoring procedure. 
	
	{\bf DeepSleepNet \cite{Supratak2017}}: Raw signals are used as input to the network. The epoch encoder is composed of two parallel 1D CNN subnetworks. The CNN layers in the subnetworks are designed to have different kernel sizes and pooling factors in order to learn features at different resolutions. The features learned by the two subnetworks are concatenated before presented to the sequence encoder. The sequence encoder, on the other hand, is implemented by two LSTM-based bidirectional RNN layers, one situated on top of the other. Residual connections \cite{He2015} are used to combine the epoch-wise features and the sequence-wise features before classification takes place. Of note, we used the end-to-end DeepSleepNet variant presented in \cite{Phan2019a} in this study.
	
	{\bf FCNN-RNN \cite{Phan2019a}}: The network resembles DeepSleepNet in several aspects: raw-signal input, the epoch encoder's reliance on CNN, and the sequence encoder's reliance on bidirectional RNN. However, its design features several differences from DeepSleepNet. First, the epoch encoder is implemented by a single CNN which makes use of full convolution \cite{Long2015} (i.e., without explicit pooling layers). Second, only one bidirectional RNN layer is employed in the sequence encoder. Third, the residual connection is omitted. These changes help reduce the network footprint significantly, more than 4 times smaller than that of DeepSleepNet.
	
	\setlength\tabcolsep{2.25pt}
	\begin{table}[!t]
		\caption{Summary of the networks employed in the study. ``TF'' stands for \emph{time-frequency} input.}
		\vspace{-0.2cm}
		\footnotesize
		\begin{center}
			\begin{tabular}{|>{\arraybackslash}m{0.8in}|>{\centering\arraybackslash}m{0.4in}|>{\centering\arraybackslash}m{0.55in}|>{\centering\arraybackslash}m{0.55in}|>{\centering\arraybackslash}m{0.7in}|>{\centering\arraybackslash}m{0in} @{}m{0pt}@{}}
				\cline{1-5}
				Network & Input & \makecell{Epoch\\Encoder} & \makecell{Sequence\\Encoder} & \makecell{\#parameters\\ (single-channel)} & \parbox{0pt}{\rule{0pt}{0ex+\baselineskip}} \\ [0ex]  	
				
				\cline{1-5}				
				SeqSleepNet & TF & RNN  & RNN & $1.6\times 10^5$  & \parbox{0pt}{\rule{0pt}{0.5ex+\baselineskip}} \\ [0ex]  	
				SleepTransformer & TF & Transformer & Transformer  & $3.7\times 10^6$ & \parbox{0pt}{\rule{0pt}{0.5ex+\baselineskip}} \\ [0ex]  	
				
				DeepSleepNet & Raw & CNN & RNN & $2.3 \times 10^7$& \parbox{0pt}{\rule{0pt}{0.5ex+\baselineskip}} \\ [0ex]  	
				FCNN-RNN & Raw & CNN & RNN & $5.6\times 10^6$ & \parbox{0pt}{\rule{0pt}{0.5ex+\baselineskip}} \\ [0ex]  	
				
				XSleepNet1& TF+Raw & RNN+CNN & RNN  & $5.7\times10^6$ & \parbox{0pt}{\rule{0pt}{0.5ex+\baselineskip}} \\ [0ex]  	
				XSleepNet2 & TF+Raw & RNN+CNN & RNN  & $5.7\times10^6$ & \parbox{0pt}{\rule{0pt}{0.5ex+\baselineskip}} \\ [0ex]  	
				
				\cline{1-5} 
			\end{tabular}
		\end{center}
		\label{tab:networks}
	\end{table}
	
	{\bf XSleepNet1 \cite{Phan2021c}}: This is a hybrid network which, in essence, accommodates SeqSleepNet and FCNN-RNN in its two network streams, respectively. It is principally designed to leverage the complementarity of SeqSleepNet (i.e., a small network solely relying on RNN) and FCNN-RNN (i.e., a larger network relying on CNN and RNN) to gain robustness to the amount of training data. Effectively, the network receives both raw-signal and time-frequency inputs which are interpreted as multiple views of the same underlying data. A multi-view learning algorithm is devised to train the network in such a way that a good multi-view representation is obtained. To that end, the learning pace of the network streams is adapted individually according to their generalization and overfitting behavior. Specially, learning on the stream that is generalizing well is encouraged with a large weight while the one that is overfitting is discouraged with a small weight. 
	
	{\bf XSleepNet2 \cite{Phan2021c}}: This network essentially shares the same architecture as XSleepNet1. The key difference between the two networks is in their multi-view learning algorithms. XSleepNet2 relies on a second-order approximation (i.e., tangents of the loss curves) to compute the adapting weights for the network streams whereas XSleepNet1 uses a first-order approximation (i.e., spontaneous values of the losses) for this purpose. Interested readers are encouraged to refer to \cite{Phan2021c} for more details.
	
	\section{Ensemble Methods}
	\label{sec:ensemble}
	
	Ensemble \cite{Dietterich2000} is a well-established machine learning approach to construct a committee model by combining existing learned ones. In general, an ensemble model often offers better performance than its individual base models. In fact, ensemble models were found to work well for automatic sleep staging with more conventional machine learning algorithms, such as Support Vector Machines \cite{Koley2012,Alickovic2018}. However, they are rarely considered as a core building block for the task in the deep learning era although a few positive results were reported, for example, in \cite{Phan2019d} which combined SeqSleepNet \cite{Phan2019a} and DeepSleepNet \cite{Supratak2017} and in \cite{Perslev2021} which fused model instances trained with different channels of the same database. 
	
	Here, we revisit the ensemble approach and form two ensemble models leveraging the six base sleep stagers described in Section \ref{sec:methods} as the base models. Theoretically, for an ensemble model to be effective the base models should be highly accurate and diversified \cite{Tsymbal2005}. The six base sleep stagers meet these criteria given their diverging characteristics on the input types and/or network architectures, except for XSleepNet1 and  XSleepNet2, and their good performance (see Section \ref{ssec:experimental_results}). Two methods are employed to combine the base sleep stagers as described below.
	
	\subsection{Ensemble via averaging probability outputs}
	
	As a deep neural network, each of the base sleep stagers produces one vector of five probability values for a 30-second epoch. These probability values indicate the likelihood that the epoch is classified as one of the five sleep stages W, N1, N2, N3, and REM. A typical method to combine the base sleep stagers is to take the average of their probability outputs. Let $\mathbf{P}^m = (P^m_1, P^m_2, \ldots, P^m_C)$, where $\sum_{1 \le c \le C} P^m_c =1$, denote the vector of probability values outputted by a model $m$  where $m \in $ \{XSleepNet1, XSleepNet2, SeqSleepNet, FCNN+RNN, DeepSleepNet, SleepTransformer\} and $C=5$ is the number of sleep stages. The vector of probability values of the ensemble model is given by $\mathbf{\bar{P}} = (\bar{P}_1, \bar{P}_2, \ldots, \bar{P}_C)$, where 
	\begin{align}
		\bar{P}_c = \frac{1}{M}\sum_{m=1}^{M} \bar{P}^m_c \text{~~~for~~~} 1 \le c \le C. 
	\end{align}
	Here, $M=6$ is the number of the base sleep stagers. The predicted sleep stage $\hat{y}$ is then determined as
	\begin{align}
		\hat{y} = \argmax_c \bar{P}_c. 
	\end{align}
	
	\subsection{Ensemble via a CNN super learner}
	\label{ssec:cnn_ensemble}
\begin{figure} [!t]
		\centering
		\includegraphics[width=0.9\linewidth]{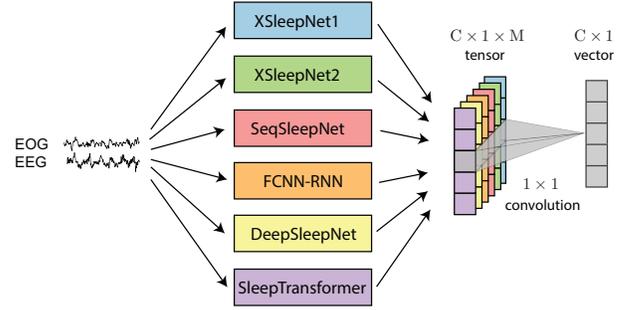}
		\caption{Fusion via the CNN with $1\times 1$ convolution.}
		\label{fig:fusion_cnn}
	\end{figure}

	Ensemble via averaging probability outputs of the base models attributes the base models equally with equal weights of $\frac{1}{M}$. Alternatively, the individual weights associated with the base models can be learned from data. Inspired by the idea of \emph{Super Learner} presented in \cite{Ju2017}, we propose a simple CNN with $1\!\times\!1$ convolution for this purpose, as illustrated in Figure \ref{fig:fusion_cnn}. Given a 30-second epoch, the probability outputs from the base models are stacked to form an tensor of size $C\!\times\!1\!\times\!M$ which will be fed into the CNN as input. With this configuration, $M$ acts as the channel dimension of the input tensor. The CNN architecture is composed of a single convolutional layer with a single $1\!\times\!1$ kernel. Convolving the kernel with the input tensor produces an output vector of size $C\!\times\!1$ which is then passed through softmax activation to translate it into a vector of probabilities values. The $1\!\times\!1$ kernel has exactly $M$ parameters, one for each of the base models, and will be learned via network training. The network is trained to minimize the cross-entropy loss. 
	
	In the experiments, it is of paramount importance that the CNN-based super learner was trained using a validation set that was set aside for model selection (see Section \ref{sssec:training}) rather than the training set. The rational is that the validation set was not directly used for training the base models, thus, avoiding overfitting the training data at this stage.
	
	\section{Experiments}
	\label{sec:experiments}
	
	\subsection{Experimental design}
	\subsubsection{Training}
	\label{sssec:training}
	The Baseline data subset was employed as the training data. Of note, 10\% (44 subjects) of the training data were left out as the validation set for early stopping purpose. Each of the sleep staging networks was trained using the training data for 10 training epochs. During the training course, the networks were validated on the validation data every 100 training steps and early stopping was activated after 100 evaluations without accuracy improvement on the validation data.
	
	In particular, the CNN super learner used for ensemble described in Section \ref{ssec:cnn_ensemble} was trained using the aforementioned validation set for 100 epochs with early stopping. 
	
	\subsubsection{Testing}
	The two subsets, Follow-up and Non-randomized, were used as two test data subsets separately. It is worth re-iterating that the former mostly consists of the same children as in the training data (i.e., the Baseline subset) while the subjects in the latter are completely new. One may be concerned about data leakage of the Follow-up test subset; however, even though the Baseline and Follow-up subsets are originated from the same subjects, mismatch in their distributions is expected given that the latter were collected 7 months after the former and after clinical intervention. Evaluating the trained networks on this test subset will shed light on the networks' performance under concept drift \cite{Tsymbal2004}. At the same time, assessing the networks' performance on the Non-randomized test subset will prove their generalization on completely new subjects. 
	
	\subsubsection{Network's initialization}
	Similar to many other domains, a large training database (i.e., thousands of subjects \cite{Phan2021c, Olesen2021, Biswal2018a, Guillot2021}) has been proven to improve generalization of deep neural networks for automatic sleep staging. Thus, the training data (440 subjects) in this study is still arguably small. Inspired by \cite{Phan2021a}, in addition to random initialization, we also investigated pretraining as an alternative approach for network initialization. That is, a network was firstly pretrained with a large external database and the pretrained network was afterwards utilized as the starting point to be further trained (i.e., finetuned) on the training data. This approach has been shown to be effective in mitigating overfitting, and hence, improving generalization, particularly when the training data is small. Here, C4-A1 EEG and ROC-LOC EOG extracted from the SHHS database (5,791 subjects) \cite{Zhang2018,Quan1997} was employed for pretraining purpose.
	
	\subsection{Parameters and metrics}
	The networks employed in this study were configured as in their original works. We also followed the procedures used in the original works in to extract the time-frequency input (i.e., the logarithmic magnitude spectrogram) when needed. The implementation was based on the \emph{Tensorflow} framework \cite{Abadi2016}. 
	
	We used the overall metrics, including accuracy, macro F1-score (MF1), Cohen's kappa ($\kappa$) \cite{McHugh2012}, sensitivity, and specificity to assess the automatic sleep staging performance. 
	
	Quantifying predictive uncertainty of the models is also important for clinical use \cite{Phan2021b} as those data epochs predicted with high uncertainty can be deferred to sleep experts for further manual inspection \cite{Becker2021}. In this regard, a model with low predictive uncertainty is preferable as it will reliably defer less data epochs to be manually checked. We used two metrics to evaluate the predictive uncertainty of a network: the negative log-likelihood (NLL) \cite{Quinonero-Candela2006} and the Brier score (BS) \cite{Brier1950}. Given a data epoch $\mathbf{S}$ with the groundtruth sleep stage $y$, NLL and BS are defined as in (\ref{eq:nll}) and (\ref{eq:brierscore}), respectively: 
	\begin{align}
		NLL &= -\sum\nolimits_{c=1}^C y_c\log\hat{y}_c, \label{eq:nll} \\
		BS &= \frac{1}{C}\sum\nolimits_{c=1}^C(y_c - \hat{y}_c)^2. \label{eq:brierscore}
	\end{align}
	In above equations, $y_c = 1$ if $y = c$, and 0 otherwise. $\hat{y}_c = P(c\,|\,\mathbf{S})$ is the probability of the epoch $\mathbf{S}$ being predicted as class $c$ by the model. The lower NLL and BS are, the lower predictive uncertainty the model has.
	
	\subsection{Experimental results}
	\label{ssec:experimental_results}
	
	\subsubsection{Sleep staging performance}
	Table \ref{tab:overall_performance} shows the performance obtained by the networks as well as their ensembles under both random initialization and pretraining initialization schemes. These results unravel several important points about pediatric automatic sleep staging.
	
	First, across different individual models, the obtained performance on pediatric sleep staging is comparable to adult sleep staging. For example, in case of random initialization and the Non-randomized subset, $\kappa$ of 0.828 and 0.842 obtained by XSleepNet1 with the single- and dual-channel input, respectively, are very similar to the state-of-the-art $\kappa$ reported on SHHS \cite{Phan2021c} consisting of 5,791 adults. These $\kappa$ values are even noticeably better than those reported on other popular adult PSG databases with smaller sizes, such as MASS \cite{Oreilly2014} and Sleep-EDF Expanded \cite{Kemp2000,Goldberger2000}. The relative performance between the networks also conform to that reported on adult PSG data, for example in \cite{Phan2021c}, where the multi-view XSleepNet1 and XSleepNet2 consistently outperformed the single-view counterparts across the test subsets, the initialization schemes, and the number of channels used. For instance, with random initialization and single-channel EEG, XSleepNet1 improved the overall accuracy by 0.8\% absolute over the best single-view networks, SeqSleepNet and FCNN+RNN, on the Follow-up and Non-randomized subset, respectively. These accuracy gaps became much narrower with the use of dual-channel EEG$\cdot$EOG (i.e., reduced to 0.2\% and 0.5\% absolute) or pretraining initialization (i.e., reduced to 0.5\% and 0.1\% absolute) or both (i.e., reduced to 0.2\% and 0.3\% absolute). On the other hand, SleepTransformer appeared to underperform other counterparts under the random initialization regime, most likely due to Transformer's data-hungry nature and the small size of the training data, as similarly observed in \cite{Phan2021b}. This is also supported by the observation that it performed comparably to the other single-view competitors after being pretrained beforehand with the large SHHS database.
	
	Second, both the ensemble models consistently resulted in better performance than all the individual models. Take random initialization and the Non-randomized test subset for example, Average Ensemble improved the overall accuracy by $1.1\%$ and $1.2\%$ absolute compared to the average overall accuracy of the six individual models. This was not only observed on the overall metrics but also over most sleep stages as evidenced by the class-wise MF1 in Table \ref{tab:classwise_performance}. However, negligible difference was seen from the performance of the two ensemble methods. In other words, the advanced ensemble method with CNN is not necessarily better than the simple averaging strategy in term of performance whilst it requires additional training.
	
	\setlength\tabcolsep{1.65pt}
	\begin{table*}[!t]
		\caption{Overall performance obtained by the individual networks and their ensemble models.}
		\scriptsize
		\vspace{-0.2cm}
		\begin{center}
			\begin{tabular}{|>{\centering\arraybackslash}m{0.45in}|>{\arraybackslash}m{0.7in}||>{\centering\arraybackslash}m{0.225in}|>{\centering\arraybackslash}m{0.275in}|>{\centering\arraybackslash}m{0.225in}|>{\centering\arraybackslash}m{0.225in}|>{\centering\arraybackslash}m{0.225in}||>{\centering\arraybackslash}m{0.225in}|>{\centering\arraybackslash}m{0.275in}|>{\centering\arraybackslash}m{0.225in}|>{\centering\arraybackslash}m{0.225in}|>{\centering\arraybackslash}m{0.225in}||>{\centering\arraybackslash}m{0.225in}|>{\centering\arraybackslash}m{0.275in}|>{\centering\arraybackslash}m{0.225in}|>{\centering\arraybackslash}m{0.225in}|>{\centering\arraybackslash}m{0.225in}||>{\centering\arraybackslash}m{0.225in}|>{\centering\arraybackslash}m{0.275in}|>{\centering\arraybackslash}m{0.225in}|>{\centering\arraybackslash}m{0.225in}|>{\centering\arraybackslash}m{0.225in}|>{\centering\arraybackslash}m{0in} @{}m{0pt}@{}}
				
				\cline{1-22}
				\multirow{3}{*}{\makecell{Subset}} &  \multirow{3}{*}{\makecell{System}} &  \multicolumn{10}{c||}{Random initialization} & \multicolumn{10}{c|}{Pretraining initialization} & \parbox{0pt}{\rule{0pt}{1ex+\baselineskip}} \\ [0ex]  	
				\cline{3-22}
				
				& & \multicolumn{5}{c|}{EEG} & \multicolumn{5}{c||}{EEG$\cdot$EOG} & \multicolumn{5}{c|}{EEG} & \multicolumn{5}{c|}{EEG$\cdot$EOG} & \parbox{0pt}{\rule{0pt}{1ex+\baselineskip}} \\ [0ex]  	
				\cline{3-22}
				& & Acc. & $\kappa$ & MF1 & Sens. & Spec. & Acc. & $\kappa$ & MF1 & Sens. & Spec. & Acc. & $\kappa$ & MF1 & Sens. & Spec. & Acc. & $\kappa$ & MF1 & Sens. & Spec. & \parbox{0pt}{\rule{0pt}{1ex+\baselineskip}} \\ [0ex]  	
				
				\cline{1-22}
				\multirow{8}{*}{\makecell{Follow-up}}   & Average Ens. & $\bm{88.9}$ & $\bm{0.852}$ & $\bm{85.3}$ & ${85.2}$ & $\bm{97.1}$ & $\bm{89.8}$ & $\bm{0.864}$ & $\bm{86.7}$ & $\bm{86.6}$ & $\bm{97.3}$ & $\bm{89.2}$ & $\bm{0.857}$ & $\bm{85.7}$ & $\bm{85.5}$ & $\bm{97.1}$ & $\bm{90.0}$ & $\bm{0.867}$ & $\bm{87.0}$ & $\bm{86.9}$ & $\bm{97.3}$ & \parbox{0pt}{\rule{0pt}{0ex+\baselineskip}} \\ [0ex]  	
				
				& CNN-based Ens. & $\bm{88.9}$ & $0.852$ & $\bm{85.3}$ & $\bm{85.2}$ & $\bm{97.1}$ & $\bm{89.8}$ & $\bm{0.864}$ & $\bm{86.7}$ & $\bm{86.5}$ & $\bm{97.3}$ & $\bm{89.2}$ & $0.856$ & $85.6$ & $\bm{85.5}$ & $\bm{97.1}$ & $\bm{90.0}$ & $\bm{0.867}$ & $86.9$ & $86.8$ & $\bm{97.3}$ & \parbox{0pt}{\rule{0pt}{0ex+\baselineskip}} \\ [0ex]  	
				
				&   XSleepNet1 &   ${88.6}$ & ${0.849}$ & ${85.2}$ & $\bm{85.3}$ & ${97.0}$ & $89.2$ & ${0.857}$ & ${86.1}$ & $\bm{86.5}$ & ${97.1}$ & ${88.7}$ & ${0.849}$ & $84.9$ & $84.9$ & ${97.0}$ &  $89.3$ & $0.858$ & ${86.4}$ & $\bm{86.9}$ & ${97.2}$ &  \parbox{0pt}{\rule{0pt}{0ex+\baselineskip}} \\ [0ex]  	
				
				&   XSleepNet2  &  $88.3$ & $0.844$ & $84.8$ & $84.8$ & $96.9$ &  ${89.3}$ & ${0.857}$ & $86.0$ & $85.8$ & ${97.1}$ & $88.6$ & ${0.849}$ & ${85.0}$ & ${85.2}$ & ${97.0}$ &  ${89.4}$ & ${0.859}$ & ${86.4}$ & $86.6$ & ${97.2}$ & \parbox{0pt}{\rule{0pt}{0ex+\baselineskip}} \\ [0ex]  	
				
				&   SeqSleepNet  & $87.8$ & $0.838$ & $83.9$ & $83.8$ & $96.8$ & $89.0$ & $0.854$ & $86.0$ & $86.0$ & $97.1$ &  $88.4$ & $0.846$ & $84.9$ & $85.2$ & $96.9$ & $89.0$ & $0.853$ & $86.0$ & $86.5$ & $97.1$ & \parbox{0pt}{\rule{0pt}{0ex+\baselineskip}} \\ [0ex]  	
				
				&   DeepSleepNet  &  $87.6$ & $0.835$ & $84.0$ & $84.8$ & $96.8$ & $88.6$ & $0.848$ & $85.3$ & $85.7$ & $97.0$ & $88.3$ & $0.844$ & $84.6$ & $84.6$ & $96.9$ & $88.5$ & $0.847$ & $85.5$ & $86.0$ & $97.0$ &  \parbox{0pt}{\rule{0pt}{0ex+\baselineskip}} \\ [0ex]  	
				
				&   FCNN+RNN   & $87.8$ & $0.837$ & $83.7$ & $83.6$ & $96.8$ & $88.5$ & $0.847$ & $85.2$ & $85.2$ & $96.9$ & $88.0$ & $0.840$ & $84.1$ & $84.2$ & $96.8$ & $87.7$ & $0.847$ & $83.6$ & $82.8$ & $96.7$ & \parbox{0pt}{\rule{0pt}{0ex+\baselineskip}} \\ [0ex]  	
				
				&   SleepTransformer &  $86.9$ & $0.825$ & $81.5$ & $81.2$ & $96.5$ & $88.2$ & $0.842$ & $83.8$ & $83.2$ & $96.8$ & $88.3$ & $0.843$ & $83.9$ & $82.8$ & $96.8$ & $89.1$ & $0.854$ & $85.2$ & $84.4$ & $97.0$ &  \parbox{0pt}{\rule{0pt}{0ex+\baselineskip}} \\ [0ex]  	
				
				\cline{1-22}
				
				\multirow{8}{*}{\makecell{Non-\\randomized}}   & Average Ens. & $\bm{87.4}$ & $\bm{0.833}$ & $\bm{83.8}$ & $\bm{83.5}$ & $\bm{96.7}$ & $\bm{88.6}$ & $\bm{0.849}$ & $\bm{85.5}$ & $\bm{85.1}$ & $\bm{96.9}$ & $\bm{87.7}$ & $\bm{0.837}$ & $\bm{84.2}$ & $\bm{83.7}$ & $\bm{96.7}$ & $\bm{88.8}$ & $\bm{0.852}$ & $\bm{85.8}$ & $\bm{85.4}$ & $\bm{97.0}$ & \parbox{0pt}{\rule{0pt}{0ex+\baselineskip}} \\ [0ex]  	

				& CNN-based Ens. & $\bm{87.4}$ & $\bm{0.833}$ & $\bm{83.8}$ & ${83.4}$ & $\bm{96.7}$ & $\bm{88.6}$ & $\bm{0.849}$ & $\bm{85.5}$ & ${85.0}$ & $\bm{96.9}$ & $87.6$ & $0.836$ & $84.1$ & $\bm{83.7}$ & $\bm{96.7}$ & $\bm{88.8}$ & $\bm{0.852}$ & $\bm{85.8}$ & $\bm{85.4}$ & $\bm{97.0}$ & \parbox{0pt}{\rule{0pt}{0ex+\baselineskip}} \\ [0ex]  	

				&   XSleepNet1  &  ${87.0}$ & ${0.828}$ & ${83.6}$ & ${83.4}$ & ${96.6}$ &  ${88.0}$ & ${0.842}$ & ${85.0}$ & $\bm{85.1}$ & ${96.8}$ & $87.0$ & $0.828$ & $83.3$ & $83.0$ & ${96.6}$ &  $88.1$ & $0.843$ & ${85.2}$ & $\bm{85.4}$ & ${96.9}$ & \parbox{0pt}{\rule{0pt}{0ex+\baselineskip}} \\ [0ex]  	
				
				&  XSleepNet2 &  $86.9$ & $0.826$ & $83.4$ & $83.1$ & $96.5$ &  ${88.0}$ & $0.841$ & $84.7$ & $84.2$ & ${96.8}$ & $87.0$ & ${0.829}$ & ${83.5}$ & ${83.5}$ & ${96.6}$ & ${88.2}$ & ${0.844}$ & $85.1$ & $85.1$ & ${96.9}$ & \parbox{0pt}{\rule{0pt}{0ex+\baselineskip}} \\ [0ex]  	
				
				&   SeqSleepNet & $86.2$ & $0.818$ & $82.3$ & $82.0$ & $96.4$ & $87.5$ & $0.835$ & $84.5$ & $84.2$ & $96.7$ & $86.9$ & $0.826$ & $83.4$ & $83.3$ & $96.5$ & $87.7$ & $0.838$ & $84.9$ & $85.1$ & $96.7$ & \parbox{0pt}{\rule{0pt}{0ex+\baselineskip}} \\ [0ex]  	
				
				&   DeepSleepNet  &  $86.0$ & $0.815$ & $82.4$ & $82.7$ & $96.3$ & $87.1$ & $0.829$ & $83.8$ & $83.8$ & $96.6$  & $86.8$ & $0.825$ & $83.2$ & $82.9$ & $96.5$ & $87.5$ & $0.835$ & $84.5$ & $84.7$ & $96.7$ & \parbox{0pt}{\rule{0pt}{0ex+\baselineskip}} \\ [0ex]  	
				
				&   FCNN+RNN  &  $86.3$ & $0.818$ & $82.4$ & $82.1$ & $96.4$ & $87.0$ & $0.827$ & $83.6$ & $83.3$ & $96.5$ & $86.2$ & $0.818$ & $82.5$ & $82.4$ & $96.4$ & $86.2$ & $0.816$ & $82.0$ & $81.1$ & $96.3$ & \parbox{0pt}{\rule{0pt}{0ex+\baselineskip}} \\ [0ex]  	
				
				&   SleepTransformer  &  $85.4$ & $0.806$ & $80.0$ & $79.7$ & $96.1$ & $87.0$ & $0.828$ & $82.7$ & $82.0$ & $96.5$ & $86.7$ & $0.822$ & $82.3$ & $81.1$ & $96.4$ & $87.8$ & $0.838$ & $84.0$ & $83.0$ & $96.7$ & \parbox{0pt}{\rule{0pt}{0ex+\baselineskip}} \\ [0ex]  	
				\cline{1-22}

				\cline{1-22}
			\end{tabular}
		\end{center}
		\label{tab:overall_performance}
		\vspace{-0.3cm}
	\end{table*}
	\begin{figure*} [!h]
		\centering
		\includegraphics[width=1\linewidth]{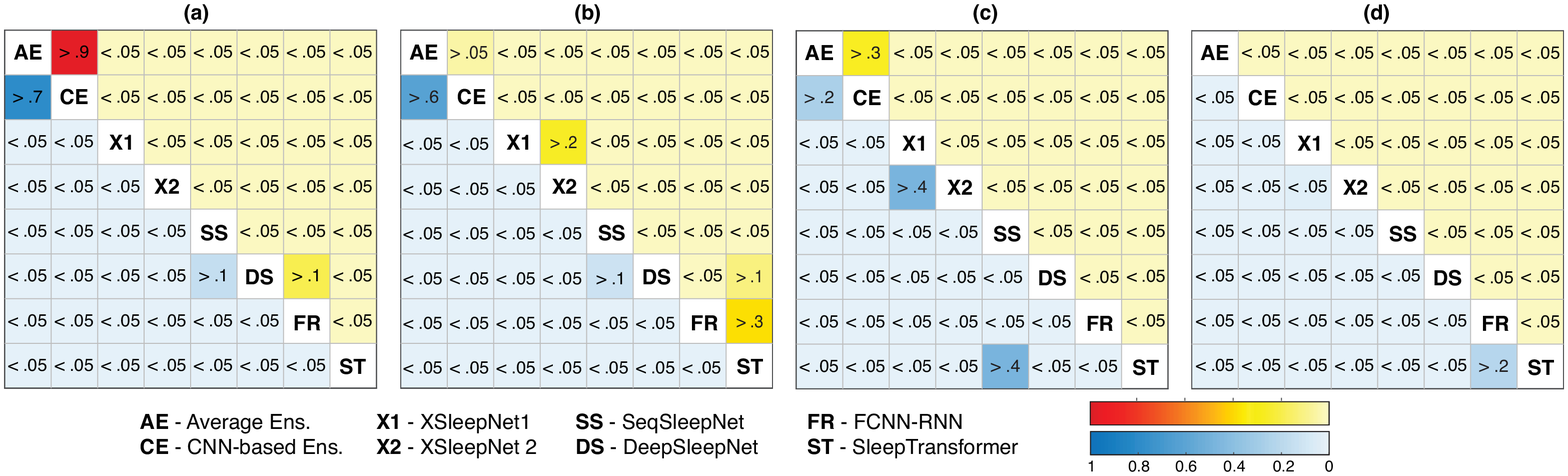}
		\caption{$p$-values obtained by the McNemar’s test on the performance improvement between different pairs of classifiers. The results in the upper and lower triangle correspond to the EEG and EEG$\cdot$EOG input, respectively. (a) Random initialization - Follow-up; (b) Random initialization - Non-randomized; (c) Pretraining initialization - Follow-up; (d) Pretraining initialization - Non-randomized.}
		\label{fig:statistical_significance}
		\vspace{-0.3cm}
	\end{figure*}

	Third, pretraining has positive effects on performance, similar to adult sleep staging \cite{Phan2021a,Guillot2021}. Between random and pretraining initialization, the latter resulted in accuracy improvement in most of the standalone models as well as the two ensemble models. However, the gains were mostly marginal since overfitting was expected to be minor given that the training data consists of hundreds of subjects. SleepTransformer was the largest beneficiary, gaining $1.3$-$1.4\%$ absolute and $0.8$-$0.9\%$ absolute on the overall accuracy with EEG and EEG$\cdot$EOG input, respectively. FCNN-RNN was the only exception which experienced accuracy drop by $0.8\%$ absolute on both the Follow-up and Non-randomized subsets when dual-channel EEG$\cdot$EOG was used. 
	
	Fourth, the obtained performances on the Follow-up subset were consistently better than those on the Non-randomized subset. For example, in case of random normalization, XSleepNet1 resulted in overall accuracies of 88.6\% and 89.2\% on the former when EEG and EEG$\cdot$EOG were used, respectively. These results were 1.6\% and 1.2\% higher than those obtained on the latter. The respective gaps were similar, 1.7\% and 1.2\%, in case of pretraining initialization. Similar patterns were also seen with the ensemble models. These results simply reflect the fact that the Follow-up subset stems from the same subjects as the  Baseline subset used for training whereas the Non-randomized subjects were completely new to the models. All in all, the results on the Non-randomized subset confirm that the automatic sleep stagers generalize to new subjects while the results on the Follow-up subset suggest that the automatic sleep stagers are robust to the concept drift given that the test data were recorded 7 months apart from the training data and after clinical intervention. 
	
	We further carried out the McNemar’s test \cite{McNemar1947} between the pairs of stagers. The statistical significance at a level of $0.05$ is seen across most of the pairs, except for a few cases, as shown in Figure \ref{fig:statistical_significance}. Moreover, the statistical significance of the ensemble models' improvement over other classifiers remained consistent over all cases while there is no significant difference between the performance of the two ensemble methods.
	
	\setlength\tabcolsep{1.65pt}
	\begin{table}[!t]
		\scriptsize
		\caption{Predictive uncertainty measures obtained by the networks and the ensemble models. AE - Average Ensemble, CE - CNN-based Ensemble, X1 - XSleepNet1, X2 - XSleepNet2, SS - SeqSleepNet, DS - DeepSleepNet, FR - FCNN-RNN, ST - SleepTransformer.}
		\vspace{-0.2cm}
		\begin{center}
			\begin{tabular}{|>{\centering\arraybackslash}m{0.45in}|>{\centering\arraybackslash}m{0.275in}|>{\centering\arraybackslash}m{0.275in}|>{\centering\arraybackslash}m{0.275in}|>{\centering\arraybackslash}m{0.275in}|>{\centering\arraybackslash}m{0.275in}||>{\centering\arraybackslash}m{0.275in}|>{\centering\arraybackslash}m{0.275in}|>{\centering\arraybackslash}m{0.275in}|>{\centering\arraybackslash}m{0.275in}|>{\centering\arraybackslash}m{0in} @{}m{0pt}@{}}
				\cline{1-10}
				\multirow{3}{*}{\makecell{Subset}} & \multirow{3}{*}{\makecell{System}} &  \multicolumn{4}{c||}{Random initialization} & \multicolumn{4}{c|}{Pretraining initialization} & \parbox{0pt}{\rule{0pt}{1ex+\baselineskip}} \\ [0ex]  	
				\cline{3-10}
				& & \multicolumn{2}{c|}{EEG} & \multicolumn{2}{c||}{EEG$\cdot$EOG} & \multicolumn{2}{c|}{EEG} & \multicolumn{2}{c|}{EEG$\cdot$EOG} & \parbox{0pt}{\rule{0pt}{1ex+\baselineskip}} \\ [0ex]  	
				\cline{3-10}
				& & NLL & BS  & NLL & BS &  NLL & BS &  NLL & BS &  \parbox{0pt}{\rule{0pt}{1ex+\baselineskip}} \\ [0ex]  	
				
				\cline{1-10}

				\multirow{8}{*}{\makecell{Follow-up}}  & AE & $\bm{0.292}$ & $\bm{0.161}$ & $\bm{0.264}$ & $\bm{0.147}$ & $\bm{0.283}$ & $\bm{0.156}$ & $\bm{0.263}$ & $\bm{0.146}$ &  \parbox{0pt}{\rule{0pt}{0ex+\baselineskip}} \\ [0ex]  	
				
				& CE & $0.343$ & $0.166$ & $0.310$ & $0.154$ & $0.336$ & $0.162$ & $0.304$ & $0.151$ &  \parbox{0pt}{\rule{0pt}{0ex+\baselineskip}} \\ [0ex]  	
				
				& X1 & $0.296$ & $0.164$ & $0.277$ & $0.155$ & $0.296$ & $0.163$ & $0.273$ & $0.153$ &  \parbox{0pt}{\rule{0pt}{0ex+\baselineskip}} \\ [0ex]  	
				
				& X2  & $0.307$ & $0.169$ & $0.274$ & $0.153$ & $0.300$ & $0.165$ & $0.272$ & $0.152$ &  \parbox{0pt}{\rule{0pt}{0ex+\baselineskip}} \\ [0ex]  	
				
				& SS  & $0.320$ & $0.176$ & $0.282$ & $0.158$ & $0.304$ & $0.168$ & $0.285$ & $0.158$ &  \parbox{0pt}{\rule{0pt}{0ex+\baselineskip}} \\ [0ex]  	
				
				& DS  & $0.324$ & $0.179$ & $0.294$ & $0.164$ & $0.305$ & $0.169$ & $0.297$ & $0.166$ &  \parbox{0pt}{\rule{0pt}{0ex+\baselineskip}} \\ [0ex]  	
				
				& FR   & $0.323$ & $0.177$ & $0.298$ & $0.166$ & $0.317$ & $0.174$ & $0.320$ & $0.176$ &  \parbox{0pt}{\rule{0pt}{0ex+\baselineskip}} \\ [0ex]  	
				
				& ST & $0.347$ & $0.189$ & $0.308$ & $0.169$ & $0.305$ & $0.168$ & $0.282$ & $0.157$ &  \parbox{0pt}{\rule{0pt}{0ex+\baselineskip}} \\ [0ex]  	
				
				\cline{1-10}
				
				\multirow{8}{*}{\makecell{Non-\\randomized}}   & AE & $\bm{0.331}$ & $\bm{0.182}$ & $\bm{0.294}$ & $\bm{0.165}$ & $\bm{0.321}$ & $\bm{0.178}$ & $\bm{0.290}$ & $\bm{0.162}$&  \parbox{0pt}{\rule{0pt}{0ex+\baselineskip}} \\ [0ex]  	
				
				& CE & $0.392$ & $0.190$ & $0.354$ & $0.174$ & $0.387$ & $0.187$ & $0.346$ & $0.170$ &  \parbox{0pt}{\rule{0pt}{0ex+\baselineskip}} \\ [0ex]  	
				
				& X1 & $0.344$ & $0.188$ & $0.310$ & $0.173$ & $0.341$ & $0.188$ & $0.306$ & $0.171$ &  \parbox{0pt}{\rule{0pt}{0ex+\baselineskip}} \\ [0ex]  	
				
				& X2  & $0.347$ & $0.190$ & $0.310$ & $0.173$ & $0.341$ & $0.187$ & $0.305$ & $0.170$ &  \parbox{0pt}{\rule{0pt}{0ex+\baselineskip}} \\ [0ex]  	
				
				& SS  & $0.365$ & $0.200$ & $0.321$ & $0.179$ & $0.343$ & $0.189$ & $0.317$ & $0.177$ &  \parbox{0pt}{\rule{0pt}{0ex+\baselineskip}} \\ [0ex]  	
				
				& DS  & $0.376$ & $0.203$ & $0.338$ & $0.186$ & $0.349$ & $0.192$ & $0.328$ & $0.181$ &  \parbox{0pt}{\rule{0pt}{0ex+\baselineskip}} \\ [0ex]  	
				
				& FR   & $0.374$ & $0.200$ & $0.343$ & $0.188$ & $0.371$ & $0.200$ & $0.363$ & $0.198$ &  \parbox{0pt}{\rule{0pt}{0ex+\baselineskip}} \\ [0ex]  	
				
				& ST & $0.393$ & $0.211$ & $0.344$ & $0.187$ & $0.357$ & $0.193$ & $0.318$ & $0.175$ &  \parbox{0pt}{\rule{0pt}{0ex+\baselineskip}} \\ [0ex]  	
				\cline{1-10}
			\end{tabular}
		\end{center}
		\label{tab:uncertainty}
		\vspace{-0.3cm}
	\end{table}
	
	\subsubsection{Predictive uncertainty}
	To quantify the predictive uncertainty of a model, we computed the average NLL and BS over all epochs of the test  data subsets individually. The results are summarized in Table \ref{tab:uncertainty}. Overall, among the six standalone models, XSleepNets resulted in lowest predictive uncertainty, outperforming  all others counterparts on both NLL and BS. Network pretraining also consistently resulted in reduced NLL and BS which were seen with both the standalone models and the ensemble ones. Interestingly, diverging patterns were seen between the two ensemble models and the simpler was better. On the one hand, Average Ensemble led to reduced predictive uncertainty, both NLL and BS, compared to the six base models. On the other hand, CNN-based Ensemble caused NLL to increase while no clear improvement was observed on BS. This observation suggests that, between the two studied ensemble methods, averaging the base models' probability outputs is more advantageous, leading to improved performance and reduced predictive uncertainty while being simple and avoiding the need for additional training. Thus, we retained the Average Ensemble model for further analysis and discussion hereafter.

	\subsubsection{Performance across age and clinical groups}
	
\begin{figure} [!t]
		\centering
		\includegraphics[width=1\linewidth]{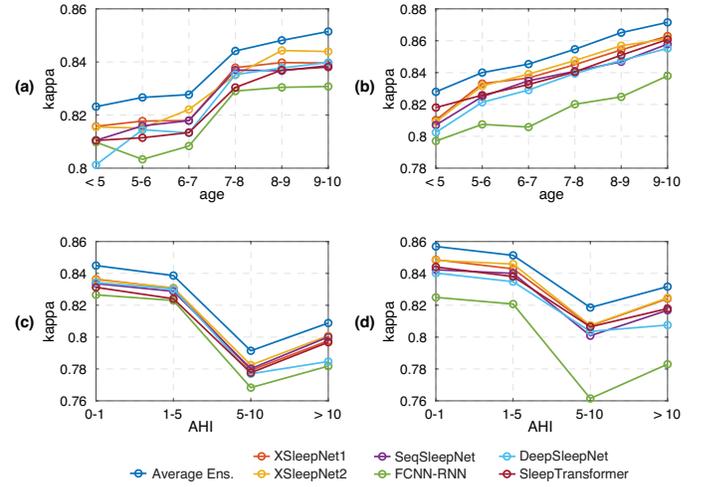}
		\caption{$k$ obtained by the models across different age and clinical (i.e. OSA severity) groups of the Non-randomized subset. (a) EEG; (b) EEG$\cdot$EOG; (c) EEG; (d) EEG$\cdot$EOG.}
		\label{fig:kappa_vs_age_ahi}
	\end{figure}

	Using the Non-randomized subset and pretraining initialization, we further investigated how the performance varies across different age and clinical groups. The results are shown in terms of $\kappa$ in Figure \ref{fig:kappa_vs_age_ahi}. On the one hand, among the models, Average Ensemble consistently stands out as the best performer regardless of ages and OSA severity while no clear winners are seen among the standalone models, particularly in case of single-channel EEG input. On the other hand, all the models exhibit increasing $\kappa$ with older groups of children. This pattern could be explained by the gradual change in sleep structure (e.g. the gradual increase of Stage 2 sleep at the expense of REM and SWS) and sleep patterns during childhood \cite{Dahl1996}. On the contrary, $\kappa$ tends to decrease with the increase of OSA severity. This downward tendency is expected due to the increase of sleep fragmentation caused by repeated occurrence of apneic arousal in OSA patients.
	
	\setlength\tabcolsep{1.75pt}
	\begin{table*}[!h]
		\caption{{Relative errors in typical clinically used measures of sleep architecture. The results are obtained from the Non-randomized subset using the pretraining initialization scheme. The lowest relative errors are printed in bold for convenience. A table cell is highlighted with color if the performance difference between the corresponding classifier and Average Ensemble is not statistically significant at a level of $0.05$.}}
		\vspace{-0.1cm}
		\footnotesize
		\begin{center}
			\begin{tabular}{|>{\arraybackslash}m{0.95in}||>{\centering\arraybackslash}m{0.55in}|>{\centering\arraybackslash}m{0.55in}|>{\centering\arraybackslash}m{0.6in}|>{\centering\arraybackslash}m{0.55in}||>{\centering\arraybackslash}m{0.55in}|>{\centering\arraybackslash}m{0.55in}|>{\centering\arraybackslash}m{0.6in}|>{\centering\arraybackslash}m{0.55in}|>{\centering\arraybackslash}m{0in} @{}m{0pt}@{}}
				\cline{1-9}
				\multicolumn{1}{|c||}{System} &  \multicolumn{4}{c||}{EEG} & \multicolumn{4}{c|}{EEG$\cdot$EOG} & \parbox{0pt}{\rule{0pt}{1ex+\baselineskip}} \\ [0ex]  	
				\cline{2-9}
				\multicolumn{1}{|c||}{}  & TST & WASO & LatREM & SE & TST & WASO & LatREM & SE & \parbox{0pt}{\rule{0pt}{1ex+\baselineskip}} \\ [0ex]  	
				
				\cline{1-9}
				
				Average Ensemble &   $8.2\pm17.8$ & $\bm{7.5\pm14.6}$ & $\bm{35.2\pm60.0}$ & $1.5\pm3.1$ & ${\bm{5.5\pm7.8}}$ & $\bm{5.3\pm7.7}$ & $\bm{25.6\pm51.1}$ & $\bm{1.0\pm1.4}$ &  \parbox{0pt}{\rule{0pt}{0ex+\baselineskip}} \\ [0ex]  	
				
				XSleepNet1  &  \cellcolor{blue!25}$8.5\pm17.4$ &  $10.1\pm21.3$& $43.0\pm69.2$ & $1.5\pm3.2$& $6.2\pm12.0$&  $6.0\pm9.8$& $34.0\pm60.7$& $1.1\pm2.1$ &  \parbox{0pt}{\rule{0pt}{0ex+\baselineskip}} \\ [0ex]  	
				
				XSleepNet2 &  \cellcolor{blue!25}${8.1\pm14.1}$& $9.8\pm19.2$& $42.7\pm65.8$& \cellcolor{blue!25}$1.5\pm2.6$ &$6.0\pm9.1$ &  \cellcolor{blue!25}$5.6\pm8.3$& $31.8\pm58.1$ & $1.1\pm1.6$ &  \parbox{0pt}{\rule{0pt}{0ex+\baselineskip}} \\ [0ex]  	

				SeqSleepNet &  \cellcolor{blue!25}$\bm{7.8\pm13.3}$ &  $8.7\pm16.7$& \cellcolor{blue!25}$36.3\pm59.8$& \cellcolor{blue!25}$\bm{1.4\pm2.4}$ &\cellcolor{blue!25}$5.8\pm9.0$& $6.3\pm11.7$& $32.7\pm60.3$ & \cellcolor{blue!25}$1.1\pm1.6$& \parbox{0pt}{\rule{0pt}{0ex+\baselineskip}} \\ [0ex]  	

				FCNN+RNN   &  $10.6\pm20.7$ & $12.2\pm22.8$& $48.4\pm71.5$& $1.9\pm3.7$&$9.0\pm17.4$ & $8.9\pm14.5$& $34.4\pm60.5$& $1.6\pm3.0$&  \parbox{0pt}{\rule{0pt}{0ex+\baselineskip}} \\ [0ex]  	
				
				DeepSleepNet &  $9.6\pm20.6$ &  $9.0\pm16.1$& $49.8\pm80.2$& $1.8\pm3.5$&$6.9\pm14.7$& $6.2\pm9.5$&  $34.9\pm60.0$& $1.2\pm2.5$&  \parbox{0pt}{\rule{0pt}{0ex+\baselineskip}} \\ [0ex]  	
				
				SleepTransformer &  $8.9\pm18.9$ &  $8.6\pm18.1$& \cellcolor{blue!25}$36.2\pm62.4$& $1.6\pm3.3$ & $6.8\pm10.7$& $7.1\pm12.9$& \cellcolor{blue!25}$26.4\pm51.0$& $1.2\pm1.9$ & \parbox{0pt}{\rule{0pt}{0ex+\baselineskip}} \\ [0ex]  	
				\cline{1-9}

			\end{tabular}
		\end{center}
		\label{tab:relative_errors}
		\vspace{-0.2cm}
	\end{table*}
	
	\subsubsection{Relative errors in clinical sleep measures}
	\label{ssec:clinical_results}

	In order to examine if Average Ensemble's performance improvement is clinically significant, we computed the relative errors in typical clinically used measures of sleep architecture, including total sleep time (TST), WASO, REM latency (LatREM), and sleep efficiency (SE), resulted by different classifiers. Again, we used the Non-randomized subset and pretraining initilization for this investigation. The results are summarized in Table \ref{tab:relative_errors}. A $t$-test was also carried to see if the difference in the relative errors between Average Ensemble and the standalone models are statistically significant. Those results which are not statistically significant at a level of $0.05$ are highlighted in color in the table. It can be seen that the Average Ensemble often led to smaller relative errors and lower variance, particularly in case of EEG$\cdot$EOG input, but the differences to the relative errors resulted by the standalone classifiers are not necessarily statistically significant in all cases. However, the fact that none of the standalone stagers is a clear winner, using the ensemble model is still favorable at the cost of increased computational overhead.

	\subsection{Further analysis and discussion}
	
	In this section, using the Non-randomized subset and the pretraining initialization scheme, we further carried out detail analyses on the agreement and the staging errors of the automatic sleep stagers to showcase that they share similar patterns on their classification errors. This observation, in turn, suggests little room for improvement in terms of staging performance, at least within the same sequence-to-sequence framework. Moreover, as indicated earlier in Section \ref{ssec:clinical_results}, these improvement would not be necessarily clinically meaningful.
	
	\subsubsection{Agreement between the automatic sleep stagers and the human scorer}
	\label{sssec:networks_agreement}
	
	\begin{figure} [!t]
		\centering
		\includegraphics[width=0.75\linewidth]{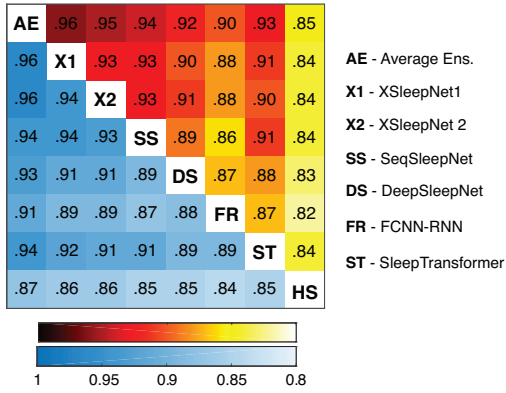}
		\caption{The agreement measured in $\kappa$ between the stagers (i.e., the studied networks, Average Ensemble, and the human scorer) on the Non-randomized subset under the pretraining initialization scheme. The results in the upper and lower triangle correspond to the EEG and EEG$\cdot$EOG input, respectively.}
		\label{fig:classifier_agreement}
		\vspace{-0.2cm}
	\end{figure}
	
	\begin{figure} [!t]
		\centering
		\includegraphics[width=1\linewidth]{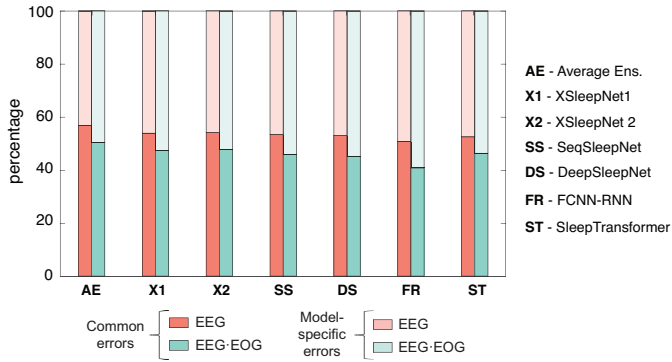}
		\caption{Percentages of the two types of errors, the common errors and others, made by the networks.}
		\label{fig:error_percentage}
		\vspace{-0.2cm}
	\end{figure}
	
	In order to elucidate the agreements among the automatic sleep stagers (i.e., the standalone models and the Average Ensemble model) and compare to the agreements between them and the human scorer, we computed $\kappa$ for all possible pairs of the stagers and show the results in Figure \ref{fig:classifier_agreement}. Of note, the pretraining initialization scheme was employed for this investigation. Overall, the agreement between every pair of the automatic stagers were considerably higher than those between them and the human scorer. The automatic stagers using the same input types (i.e., raw signal, time-frequency image, and both) tended to agree to one another more than between the stagers using different input types.  However, given the range of $\kappa$ between $0.865$ and $0.961$, the agreement level was ``almost perfect'' (according to the interpretation of Cohen's kappa \cite{Cohen1960}) and considerably higher than the ``substantial'' level between human scorers (for example, $k=0.76$ as reported in other studies \cite{DankerHopfe2009, Rosenberg2014}). Interestingly, the attribution of the base models in the ensemble model was manifested via their agreements to the Average Ensemble model. This suggests that even though the Average Ensemble model allocates equal weights to the base models, their attributions to the ensemble are inherently proportionate to their performance.
	
	\begin{figure} [!t]
		\centering
		\includegraphics[width=1\linewidth]{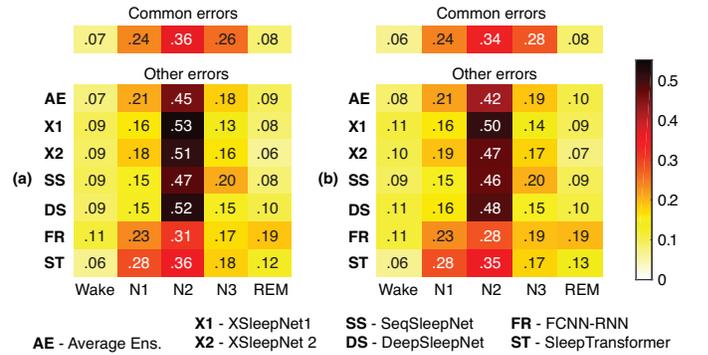}
		\caption{Distribution of the errors on made by the networks over the sleep stages. The results were obtained from the Non-randomized subset using the pretraining initialization scheme. (a) EEG;  (b) EEG$\cdot$EOG.}
		\label{fig:error_distribution}
		\vspace{-0.2cm}
	\end{figure}
	\begin{figure*} [!t]
		\centering
		\includegraphics[width=1\linewidth]{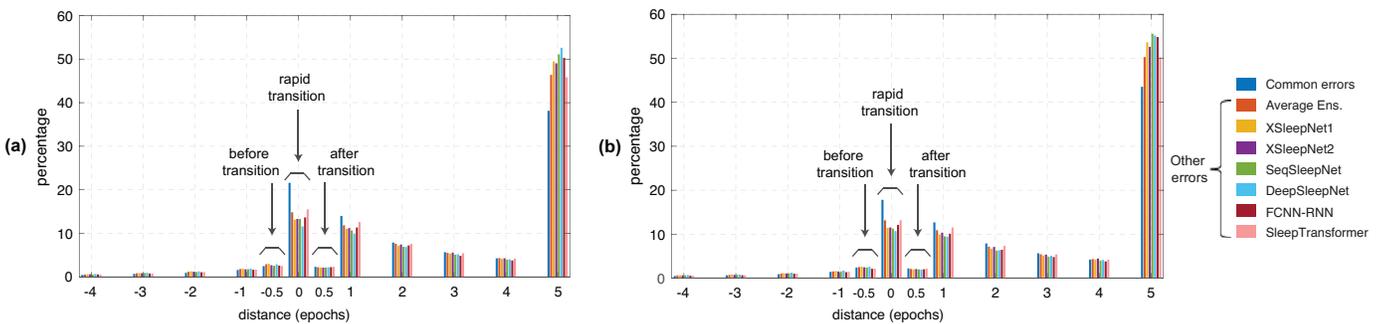}
		\caption{The distance of the misclassified epochs to their nearest transitions. In the figures, ``rapid transition'' indicates the epochs both immediately before and after the transition. The results were obtained from the Non-randomized subset and the pretraining initialization scheme. (a) EEG; (b) EEG$\cdot$EOG.}
		\label{fig:error_transition_distance}
		\vspace{-0.2cm}
	\end{figure*}
	
	\begin{figure*} [!t]
		\centering
		\includegraphics[width=0.9\linewidth]{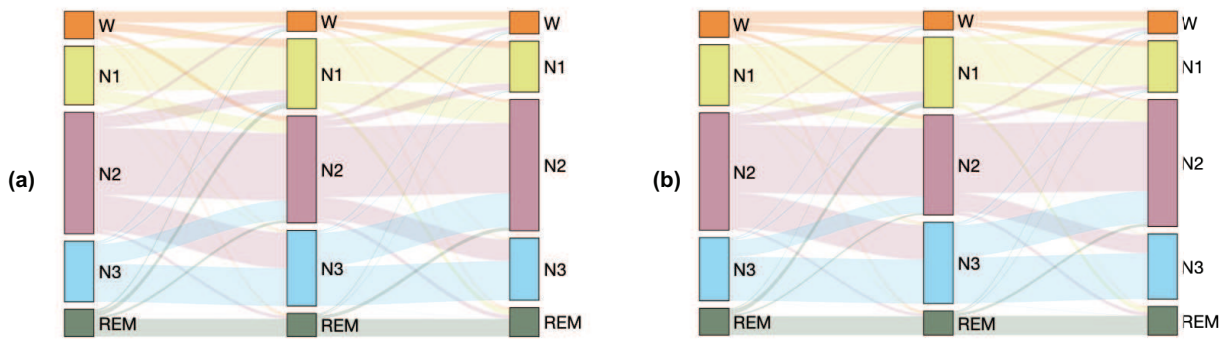}
		\caption{The stage transitions from the predecessors of the common-error epochs to themselves, and then to their successors. The results were obtained from the Non-randomized subset and the pretraining initialization scheme.  (a) EEG; (b) EEG$\cdot$EOG.}
		\label{fig:error_transitions}
		\vspace{-0.2cm}
	\end{figure*}
	
	It is also worth mentioning the study on adult PSG scoring by Mikkelsen \emph{et al.} \cite{Mikkelsen2021} which showed the higher agreement between an automatic sleep stager and a human scorer than the agreement between two human scorers themselves. Connecting this finding and our above analysis on the agreement between the automatic scorers suggests that it is probably unnecessary to create more deep learning networks for improving accuracy \emph{per se} and probably that the current automatic sleep staging algorithms are already ready for clinical use. 
	
	\subsubsection{Errors made by the automatic stagers}
	\label{sssec:networks_errors}
	
	In this section, we aim to shed some light on the errors made by the automatic stagers. To this end, we distinguish two types of errors: (1) the errors which were commonly made by all the stagers and (2) other errors. The former can be interpreted as \emph{unrecognizable} whereas the latter, in contrast, can be interpreted as \emph{recognizable} as they were correctly classified by at least one stager. 
	
	The percentages of the two types of errors are shown in Figure \ref{fig:error_percentage}. The figure reveals a significant amount of sleep epochs systematically misclassified by all the stagers, constituting more than 50\% of all the errors each of them made in case of single-channel EEG. It reduced to below 50\% when EOG was additionally included but still remained a large portion. Furthermore, most of the common errors corresponded to non-rapid eye movement (NREM) stages (i.e., N1, N2, and N3) which constituted up to about 86\%. This can be explained by the fact that N2 and N3 are the major classes in the database (cf. Table \ref{tab:databases}) while N1, as similar to adult sleep, is much less well-defined than other stages. Figure \ref{fig:error_transition_distance} further shows that more than 60\% (in case of single-channel EEG) and 55\% (in case of dual-channel EEG$\cdot$EOG) of the common errors were close to cross-stage transitions, at most 4 epochs away from their nearest cross-stage transitions. Moreover, around 20\% of them were rapid-transition epochs which are, in general, challenging to be recognized correctly as they tend to convey features of multiple sleep stages and, typically, manual labelling these epochs is highly subjective. Interestingly, the figure also reveals that, with the same nearest-to-transition distance, the networks tended to misclassify those before transitions more than those after transitions. No such a difference was seen in case of adult PSG staging when a similar analysis was conducted using SeqSleepNet in \cite{Mikkelsen2021}. We further visualize in Figure \ref{fig:error_transitions} the stage transitions from the predecessors of the common-error epochs to themselves, and then to their successors. Apparently, the compelling similarity of the transitioning patterns in the figure suggests the convergence of the common errors across different scenarios. Put simpler, there exists a set of epochs associated with some specific stage transitions that could not be recognized by the automatic sleep stagers regardless the addition of the EOG channel and the ensemble of stagers.

	Regarding other errors, as shown in Figure \ref{fig:error_distribution}, all the stagers shared a similar pattern where N2 was the most misclassified stage ($40-53\%$ by Average Ensemble, XSleepNet1, XSleepNet2, SeqSleepNet, DeepSleepNet; and $28-36\%$ by FCNN-RNN and SleepTransformer), followed by N1, and then N3. However, compared to other stagers,  FCNN-RNN and SleepTransformer appeared to have these errors distributed more evenly between N1 and N2. Regarding distances to the nearest cross-stage transitions, similar findings can be drawn from Figure \ref{fig:error_distribution} as in case of the common errors, except that the percentages of errors in the vicinity of maximally 4 epochs to the nearest transitions were lower and that the percentages of the rapid-transition epochs were also much lower. These patterns were unanimous across all the stagers.

	All in all, the above analysis confirms that the majority of the automatic sleep stagers manifested a similar pattern on their classification errors. In other words, they behaved analogously on the automatic sleep staging task. This finding is indeed complementary to the ``almost perfect'' agreement among the stagers  in Section \ref{sssec:networks_agreement}. 
	
	\section{Conclusions}
	\label{sec:conclusions}
	
	We conducted a comparative study on six different deep neural networks for automatic sleep staging on a large-scale pediatric population with a wide range of OSA severity. The benchmarking results demonstrate that the studied networks, which are the state-of-the-art algorithms in adult sleep staging, generalized well to young children, achieving an expert-level accuracy similar to that reported on adult PSGs when evaluated on new subjects. Combining the six networks into ensemble models further boosted accuracy and led to reduced predictive uncertainty. The automatic sleep stagers, the ensemble models included, were also robust to the concept drift when the test data were recorded 7 months later and after clinical intervention. However, the performance improvement did not necessarily translate into clinical significance.	Equally important, the stagers exhibited ``almost perfect'' agreement to one another and similar patterns on their classification errors. These results suggest that there is probably little room for accuracy improvement within the same sequence-to-sequence framework, if any, the improvement would be not necessarily clinically meaningful. Rather, future works should focus on entirely different concepts for automatic sleep staging and other overarching, clinician-centric challenges, such as explainability and uncertainty estimation, to accelerate clinical adoption of automatic sleep staging algorithms. Automated scoring of sleep micro-architecture such as cortical arousal, sleep spindles, and the cyclic alternating pattern is another important frontier \cite{Hartmann2019,Hartmann2020,Hartmann2021}.
	
	
	\balance
	\bibliographystyle{IEEEbib}
	\bibliography{bibliography}
	
	\appendices
	\setcounter{table}{0}
	
	\section{Class-wise MF1 obtained by the automatic sleep stagers}	
	
	\numberwithin{equation}{section}
	\numberwithin{table}{section}
	\numberwithin{figure}{section}
	
	\setlength\tabcolsep{1.65pt}
	\begin{table*}[!t]
		\caption{Class-wise MF1 obtained by the individual networks and their ensemble model.}
		\scriptsize
		\vspace{-0.2cm}
		\begin{center}
			\begin{tabular}{|>{\centering\arraybackslash}m{0.45in}|>{\arraybackslash}m{0.7in}||>{\centering\arraybackslash}m{0.225in}|>{\centering\arraybackslash}m{0.275in}|>{\centering\arraybackslash}m{0.225in}|>{\centering\arraybackslash}m{0.225in}|>{\centering\arraybackslash}m{0.225in}||>{\centering\arraybackslash}m{0.225in}|>{\centering\arraybackslash}m{0.275in}|>{\centering\arraybackslash}m{0.225in}|>{\centering\arraybackslash}m{0.225in}|>{\centering\arraybackslash}m{0.225in}||>{\centering\arraybackslash}m{0.225in}|>{\centering\arraybackslash}m{0.275in}|>{\centering\arraybackslash}m{0.225in}|>{\centering\arraybackslash}m{0.225in}|>{\centering\arraybackslash}m{0.225in}||>{\centering\arraybackslash}m{0.225in}|>{\centering\arraybackslash}m{0.275in}|>{\centering\arraybackslash}m{0.225in}|>{\centering\arraybackslash}m{0.225in}|>{\centering\arraybackslash}m{0.225in}|>{\centering\arraybackslash}m{0in} @{}m{0pt}@{}}
				\cline{1-22}
				\multirow{3}{*}{\makecell{Subset}} &  \multirow{3}{*}{\makecell{System}} &  \multicolumn{10}{c||}{Random initialization} & \multicolumn{10}{c|}{Pretraining initialization} & \parbox{0pt}{\rule{0pt}{1ex+\baselineskip}} \\ [0ex]  	
				
				&  &  \multicolumn{5}{c||}{EEG} & \multicolumn{5}{c|}{EEG$\cdot$EOG} &  \multicolumn{5}{c||}{EEG} & \multicolumn{5}{c|}{EEG$\cdot$EOG} & \parbox{0pt}{\rule{0pt}{1ex+\baselineskip}} \\ [0ex]  	
				\cline{3-22}
				& & Wake & N1 & N2 & N3 & REM & Wake & N1 & N2 & N3 & REM & Wake & N1 & N2 & N3 & REM & Wake & N1 & N2 & N3 & REM & \parbox{0pt}{\rule{0pt}{1ex+\baselineskip}} \\ [0ex]  	
				
				\cline{1-22}
				
				\multirow{8}{*}{\makecell{Follow-up}} & Average Ens. & $\bm{94.3}$ & ${63.0}$ & $\bm{88.5}$ & $\bm{91.2}$ & ${89.5}$ & $\bm{95.5}$ & $\bm{66.1}$ & $\bm{89.1}$ & $\bm{91.2}$ & $\bm{91.9}$ & $\bm{94.6}$ & $\bm{63.4}$ & $\bm{88.8}$ & $\bm{91.3}$ & $\bm{90.2}$ &  $\bm{95.6}$ & $\bm{66.5}$ & $\bm{89.1}$ & $\bm{91.4}$ & ${92.2}$ & \parbox{0pt}{\rule{0pt}{0ex+\baselineskip}} \\ [0ex]  	
				
				& CNN-based Ens. & $\bm{94.3}$ & $62.9$ & $\bm{88.5}$ & $\bm{91.2}$ & $\bm{89.6}$ & $\bm{95.6}$ & $65.7$ & $\bm{89.1}$ & $\bm{91.2}$ & $\bm{91.9}$ & $\bm{94.5}$ & $63.3$ & $\bm{88.8}$ & $\bm{91.3}$ & $\bm{90.2}$ & $\bm{95.6}$ & $66.4$ & $\bm{89.1}$ & $\bm{91.4}$ & $\bm{92.3}$ & \parbox{0pt}{\rule{0pt}{0ex+\baselineskip}} \\ [0ex]  	
				
				&   XSleepNet1 &   ${94.1}$ & $\bm{63.2}$ & ${88.1}$ & ${90.9}$ & $\bm{89.6}$ & $95.0$ & ${65.0}$ & $88.4$ & ${91.0}$ & ${91.2}$ & ${94.2}$ & $61.7$ & ${88.2}$ & $91.1$ & $89.4$ & ${95.4}$ & ${65.9}$ & $88.2$ & $91.0$ & $91.6$ & \parbox{0pt}{\rule{0pt}{0ex+\baselineskip}} \\ [0ex]  	
				
				&   XSleepNet2  &  $93.6$ & $62.8$ & $88.0$ & $90.6$ & $88.9$ & $95.0$ & $64.3$ & ${88.6}$ & ${91.0}$ & $91.1$ & $94.1$ & ${62.2}$ & $88.1$ & ${91.2}$ & $89.4$ & $95.2$ & $65.6$ & ${88.5}$ & ${91.1}$ & $91.4$ & \parbox{0pt}{\rule{0pt}{0ex+\baselineskip}} \\ [0ex]  	

				&   SeqSleepNet  & $93.6$ & $60.2$ & $87.1$ & $90.3$ & $88.5$ & ${95.4}$ & $64.9$ & $88.0$ & $90.6$ & $91.0$ &  $94.0$ & $60.1$ & $86.3$ & $90.0$ & $87.7$ & $95.0$ & $65.2$ & $88.0$ & $90.7$ & $91.2$ & \parbox{0pt}{\rule{0pt}{0ex+\baselineskip}} \\ [0ex]  	
				
				&   DeepSleepNet  &  $92.9$ & $60.7$ & $87.4$ & $90.5$ & $88.5$ & $94.4$ & $63.3$ & $88.2$ & $90.2$ & $90.5$ & $93.5$ & $61.8$ & $87.8$ & $90.7$ & $89.1$ & $94.7$ & $64.7$ & $87.4$ & $90.0$ & $90.7$ & \parbox{0pt}{\rule{0pt}{0ex+\baselineskip}} \\ [0ex]  	
				
				&   FCNN+RNN   & $92.6$ & $59.5$ & $87.6$ & $90.7$ & $88.1$ & $93.8$ & $63.1$ & $88.0$ & $90.4$ & $90.7$ & $92.9$ & $60.6$ & $87.8$ & $91.0$ & $88.2$ & $93.2$ & $57.4$ & $87.2$ & $90.4$ & $89.8$ & \parbox{0pt}{\rule{0pt}{0ex+\baselineskip}} \\ [0ex]  	
				
				&   SleepTransformer &  $92.9$ & $49.8$ & $86.7$ & $90.2$ & $87.6$ & $94.9$ & $56.0$ & $87.4$ & $90.4$ & $90.4$ & $93.9$ & $57.3$ & $88.1$ & $90.4$ & ${89.7}$ & $94.7$ & $60.3$ & $88.3$ & $90.8$ & ${91.9}$ & \parbox{0pt}{\rule{0pt}{0ex+\baselineskip}} \\ [0ex]  	
				
				\cline{1-22}
				
				\multirow{8}{*}{\makecell{Non-\\randomized}} & Average Ens. &$\bm{93.5}$ & ${61.4}$ & $\bm{86.4}$ & $\bm{89.6}$ & $\bm{88.2}$ &${95.2}$ & $\bm{64.4}$ & $\bm{87.3}$ & $\bm{89.9}$ & ${90.6}$ & $\bm{93.6}$ & $\bm{62.1}$ & $\bm{86.6}$ & $\bm{89.7}$ & $\bm{88.8}$ & $\bm{95.5}$ & $\bm{65.2}$ & $\bm{87.3}$ & $\bm{90.1}$ & ${91.0}$ & \parbox{0pt}{\rule{0pt}{0ex+\baselineskip}} \\ [0ex]  	
				
				& CNN-based Ens. & $\bm{93.5}$ & $61.3$ & $\bm{86.4}$ & $\bm{89.6}$ & $\bm{88.2}$ & $\bm{95.3}$ & $64.2$ & $\bm{87.3}$ & $\bm{89.9}$ & $\bm{90.7}$ & $\bm{93.6}$ & $62.0$ & $\bm{86.6}$ & $\bm{89.7}$ & $\bm{88.8}$ & $\bm{95.5}$ & $65.0$ & $\bm{87.3}$ & $\bm{90.1}$ & $\bm{91.1}$ & \parbox{0pt}{\rule{0pt}{0ex+\baselineskip}} \\ [0ex]  	
				
				&   XSleepNet1  &  ${93.3}$ & $\bm{61.8}$ & $85.8$ & ${89.2}$ & ${87.9}$ & $94.8$ & ${63.8}$ & $86.6$ & $\bm{89.9}$ & ${89.9}$ & $93.1$ & $60.2$ & ${86.0}$ & $89.3$ & $88.0$ & ${95.2}$ & ${64.7}$ & $86.4$ & ${89.7}$ & $90.1$ &  \parbox{0pt}{\rule{0pt}{0ex+\baselineskip}} \\ [0ex]  	
				
				&  XSleepNet2 &  $93.2$ & $61.2$ & ${86.0}$ & $88.9$ & $87.7$ & $94.7$ & $62.5$ & ${86.7}$ & $89.6$ & $89.8$ & $93.4$ & $60.5$ & $85.9$ & ${89.5}$ & $88.1$ & $95.1$ & $64.1$ & ${86.6}$ & ${89.7}$ & $90.2$ & \parbox{0pt}{\rule{0pt}{0ex+\baselineskip}} \\ [0ex]  	
				
				&   SeqSleepNet & $93.1$ & $58.2$ & $84.8$ & $88.7$ & $86.7$ & ${95.0}$ & $63.2$ & $85.7$ & $89.1$ & $89.4$ &  ${93.5}$ & ${60.8}$ & $85.5$ & $89.1$ & ${88.2}$ & $95.1$ & $64.3$ & $85.9$ & $89.0$ & $90.2$ & \parbox{0pt}{\rule{0pt}{0ex+\baselineskip}} \\ [0ex]  	
				
				&   DeepSleepNet  &  $91.8$ & $59.4$ & $85.3$ & $88.8$ & $86.7$ & $93.3$ & $61.8$ & $86.3$ & $89.0$ & $88.4$  & $92.8$ & $60.7$ & $85.7$ & $89.2$ & $87.4$ & $94.5$ & $63.7$ & $85.9$ & $89.1$ & $89.6$ & \parbox{0pt}{\rule{0pt}{0ex+\baselineskip}} \\ [0ex]  	
				
				&   FCNN+RNN  &  $91.9$ & $58.7$ & $85.4$ & ${89.2}$ & $86.6$ & $92.8$ & $61.4$ & $86.1$ & $89.1$ & $88.7$ &  $91.6$ & $59.4$ & $85.4$ & $89.2$ & $86.6$ & $92.3$ & $55.8$ & $85.1$ & $88.9$ & $87.9$ & \parbox{0pt}{\rule{0pt}{0ex+\baselineskip}} \\ [0ex]  	
				
				&   SleepTransformer  &  $92.4$ & $48.2$ & $84.5$ & $88.7$ & $86.5$ & $94.9$ & $54.8$ & $85.5$ & $89.2$ & $89.1$ & $93.1$ & $55.4$ & $85.9$ & $89.0$ & $87.9$ & $94.7$ & $58.7$ & $86.2$ & $89.5$ & ${90.7}$ & \parbox{0pt}{\rule{0pt}{0ex+\baselineskip}} \\ [0ex]  	
				\cline{1-22}

				\cline{1-22}
			\end{tabular}
		\end{center}
		\label{tab:classwise_performance}
	\end{table*}

\end{document}